\documentclass[sigconf]{acmart}

\usepackage[most]{tcolorbox}
\tcbset{
  colback=gray!5!white, 
  colframe=gray!50!black, 
  colbacktitle=gray!50!black,
  coltitle=white,
  fonttitle=\bfseries\footnotesize,
  boxrule=0.5pt, 
  arc=3pt, 
  left=6pt, 
  right=6pt, 
  top=2pt, 
  bottom=2pt, 
  boxsep=4pt,
  enhanced,
}
\AtBeginDocument{%
  }

\copyrightyear{2025}
\acmYear{2025}
\acmBooktitle{The 2nd Workshop on LLMs and Generative AI for Finance at the 6th ACM International Conference on AI in Finance (ICAIF '25), November 15--18, 2025, Singapore, Singapore}

\begin{document}

\title{Resource-Efficient LLM Application for Structured Transformation of Unstructured Financial Contracts}

\author{Maruf Ahmed Mridul}
\affiliation{%
    \institution{Rensselaer Polytechnic Institute}
    \city{Troy}
    \state{New York}
    \country{USA}
    }
\email{mridum@rpi.edu}
\orcid{0009-0003-7501-4714}

\author{Oshani Seneviratne}
\affiliation{%
	\institution{Rensselaer Polytechnic Institute}
	\city{Troy}
	\state{New York}
	\country{USA}
}
\email{senevo@rpi.edu}
\orcid{0000-0001-8518-917X}

\renewcommand{\shortauthors}{Mridul and Seneviratne}

\begin{abstract}

The transformation of unstructured legal contracts into standardized, machine-readable formats is essential for automating financial workflows. The Common Domain Model (\textbf{CDM}) provides a standardized framework for this purpose, but converting complex legal documents like Credit Support Annexes (\textbf{CSAs}) into CDM representations remains a significant challenge. In this paper, we present an extension of the CDMizer framework, a template-driven solution that ensures syntactic correctness and adherence to the CDM schema during contract-to-CDM conversion. We apply this extended framework to a real-world task, comparing its performance with a benchmark developed by the International Swaps and Derivatives Association (\textbf{ISDA}) for CSA clause extraction.  Our results show that CDMizer, when integrated with a significantly smaller, open-source Large Language Model (\textbf{LLM}), achieves competitive performance in terms of accuracy and efficiency against larger, proprietary models. This work underscores the potential of resource-efficient solutions to automate legal contract transformation, offering a cost-effective and scalable approach that can meet the needs of financial institutions with constrained resources or strict data privacy requirements.
\end{abstract}



\maketitle

\section{Introduction}
The digitization and standardization of legal agreements are fundamental steps toward automating complex financial workflows, such as collateral management, regulatory reporting, and trade processing. 
In this paper, we focus on Credit Support Annexes (CSAs), which are integral components of ISDA Master Agreements that govern collateral management in over-the-counter (OTC) derivatives markets. They define critical terms such as eligible collateral, thresholds, and minimum transfer amounts, directly shaping counterparty risk exposure. However, CSAs are drafted in unstructured legal text, making them difficult to interpret consistently across institutions. This lack of standardization hinders automation, slows regulatory reporting, and increases operational risk. Digitizing CSAs into machine-readable, schema-adherent formats is therefore essential for enabling scalable collateral workflows, ensuring compliance, and improving transparency in financial markets.

The industry standard for this digital transformation is the \textbf{CDM}, a standardized, machine-readable blueprint for financial products, trades, and their lifecycle events~\cite{cdm_purpose}. Although the CDM provides a clear structure, converting unstructured legal agreements like CSAs from the over-the-counter (OTC) derivatives market into accurate CDM representations is still a major challenge. 
Significant industry efforts, including recent benchmarking work~\cite{csa_benchmark} by ISDA, are currently dedicated to addressing this complex conversion challenge.

Additionally, recent advancements in LLMs and Retrieval-Augmented Generation (RAG) offer promising avenues for automating this conversion. However, two major challenges persist, particularly for financial institutions:
\begin{itemize}
    \item \textit{Schema and Syntactical Adherence:} Direct generation by LLMs often struggles to consistently produce complex, deeply nested JSON outputs that adhere strictly to the precise CDM schema, leading to validation errors and integration issues.
    
    \item \textit{Accessibility and Cost:} Industry-leading benchmarks for contract digitization have demonstrated high accuracy predominantly through the use of very large, proprietary LLMs (e.g., GPT-4o, Claude Opus)~\cite{csa_benchmark}. These models pose significant cost, data privacy, and deployment challenges that restrict their immediate adoption by many institutions, underscoring the need for competitive, open-source alternatives.
\end{itemize}

To address the adherence challenge, we previously introduced \textbf{CDMizer}~\cite{mridul2025ai4contractsllmragpowered}, a template-driven framework that enforces syntactic correctness and schema adherence during the contract-to-CDM conversion process. CDMizer employs a recursive traversal approach, integrating RAG, to populate pre-generated, minimal CDM templates based on hierarchical context. In its initial application, CDMizer was used for converting OTC derivative contracts, demonstrating its ability to handle complex financial agreements and structure them within the CDM framework.

In this work, we build upon the CDMizer framework to tackle the accessibility and cost challenge by applying it to a critical, real-world task defined by ISDA. Specifically, we perform a rigorous comparative analysis using the \textbf{Qwen3-30B}\footnote{\url{https://huggingface.co/Qwen/Qwen3-30B-A3B-Instruct-2507}}, which is a state-of-the-art, yet significantly smaller and open-source-compatible model, against the published performance of proprietary LLMs on the ISDA CSA benchmark.

The core contributions of this paper are:
\begin{itemize}
\item \textbf{Methodology Generalization:} We validate the application of the CDMizer framework for accurately encoding ISDA CSA clauses, extending its use to a new document type beyond its initial application to OTC derivative contracts.
\item \textbf{Resource-Efficient Benchmarking:} We quantify the performance of CDMizer when integrated with the highly resource-efficient Qwen3-30B model, providing a benchmark for accessible, internal deployment in comparison to larger models.
\item \textbf{Performance vs. Accessibility Trade-offs:} We analyze the semantic accuracy gap between our resource-constrained approach and commercial State-of-the-Art (SOTA) models, highlighting a practical approach to CDM conversion that balances cost and data privacy constraints.
\end{itemize}

\section{Related Work}
The automation of legal contract transformation into standardized formats has been explored with the use of LLMs and RAG. However, most research has focused on structured contracts, leaving the conversion of unstructured financial documents, like CSAs, largely unaddressed.

Early efforts in automating OTC derivatives, such as~\cite{fries2018smart} and~\cite{armitage2022trust}, relied on structured inputs, addressing only specific tasks like automating termination procedures or transforming standardized agreements into smart contracts. These works did not explore the challenges of handling unstructured legal text in complex contracts.

In contract generation, LLMs have been applied to simpler contracts, like those in~\cite{kang2024using} and~\cite{van2023translating}, but these studies focused on domains like health insurance, which lack the complexity of financial agreements.~\cite{karanjai2024solmover} explored contract translation for blockchain using structured representations but did not tackle unstructured legal language.

The integration of RAG has proven useful in improving LLM-generated content, as demonstrated by~\cite{parvez2021retrieval}, which enhanced the accuracy and coherence of contract generation. This approach is particularly valuable for financial contracts, where extracting structured information from unstructured text is critical. However, existing evaluation methods, such as those used in~\cite{kang2024using} and~\cite{vaithilingam2022expectation}, fall short when it comes to assessing the complex and nuanced nature of financial contracts.

In our previous work~\cite{mridul2025ai4contractsllmragpowered}, we introduced the CDMizer framework to address these challenges by applying a template-driven approach to convert OTC derivative contracts into CDM representations. This method ensured syntactic correctness and schema adherence, making it effective for simpler derivative contracts.  This paper extends CDMizer by applying it to a new dataset of CSA contracts and evaluating its performance against an industry benchmark.


\section{Experimental Setup}
\begin{figure}[h]
    \centering
    \includegraphics[width=0.5\linewidth]{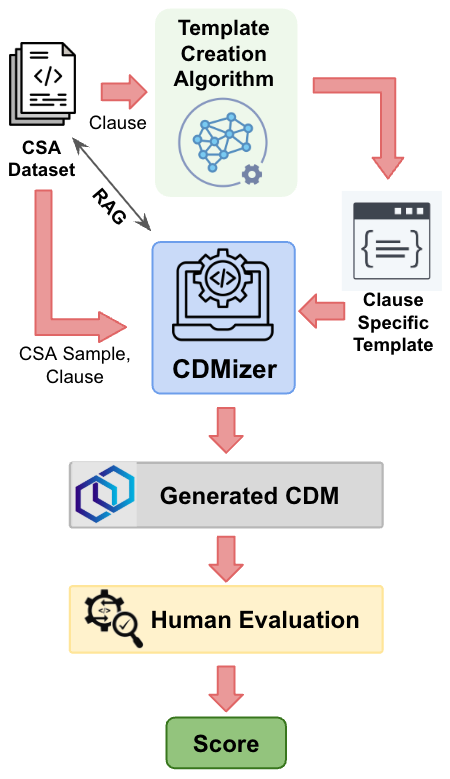}
    \caption{Overall Experimental Workflow}
    \label{fig:workflow}
\end{figure}

Our experimental approach leverages the core architecture of \textbf{CDMizer}~\cite{mridul2025ai4contractsllmragpowered} to execute the contract-to-CDM conversion task. As outlined in the workflow diagram shown in Figure ~\ref{fig:workflow}, the process follows three distinct stages - 1) creating templates specific to the target CSA clauses from the CDM schema and relevant examples, 2) populating these templates using the chosen LLM, and 3) evaluating the generated CDM outputs.

\begin{table*}[t]
\centering
\caption{Comparative scores of different models (With RAG)}
\label{tab:rag_comparison}
\begin{tabular}{lcccc}
\toprule
\textbf{Model / Method} & \textbf{Base and Eligible Currency} & \textbf{MTA} & \textbf{Threshold} & \textbf{Rounding} \\
\midrule
GPT-4o (\textasciitilde200B) & 98.00 & 93.00 & 92.00 & 98.00 \\
GPT-o1 (\textasciitilde200B) & 100.00 & 93.00 & 90.00 & 100.00 \\
Claude 3.5 Sonnet v2 (175B+) & 100.00 & 85.00 & 88.00 & 100.00 \\
Claude 3.7 Sonnet (100B+) & 100.00 & 90.00 & 90.00 & 100.00 \\
Claude 3 Opus (100B+) & 98.00 & 87.00 & 90.00 & 95.00 \\
DeepSeek R1 (671B, 37B active) & 88.00 & 83.00 & 85.00 & 98.00 \\
Nova Pro (\textasciitilde90B) & 95.00 & 72.00 & 90.00 & 95.00 \\
Llama 3.3 (70B) & 98.00 & 70.00 & 82.00 & 98.00 \\
\midrule
\textbf{CDMizer with Qwen3 (30.5B)} & \textbf{97.88} & \textbf{79.15} & \textbf{88.24} & \textbf{93.39} \\
\textbf{Rank (out of 9 models)} & \textbf{7\textsuperscript{th}} & \textbf{7\textsuperscript{th}} & \textbf{6\textsuperscript{th}} & \textbf{9\textsuperscript{th}} \\
\bottomrule
\end{tabular}
\end{table*}

\begin{table*}[t]
\centering
\caption{Comparative scores of different models (Without RAG)}
\label{tab:baseline_comparison}
\begin{tabular}{lcccc}
\toprule
\textbf{Model / Method} & \textbf{Base and Eligible Currency} & \textbf{MTA} & \textbf{Threshold} & \textbf{Rounding} \\
\midrule
GPT-4o (\textasciitilde 200B) & 98.00 & 37.00 & 87.00 & 97.00 \\
GPT-o1 (\textasciitilde200B) & 97.00 & 37.00 & 80.00 & 98.00 \\
Claude 3.5 Sonnet v2 (175B+) & 98.00 & 30.00 & 40.00 & 98.00 \\
Claude 3.7 Sonnet (100B+) & 98.00 & 38.00 & 40.00 & 90.00 \\
Claude 3 Opus (100B+) & 95.00 & 35.00 & 88.00 & 92.00 \\
DeepSeek R1 (671B, 37B active) & 85.00 & 37.00 & 68.00 & 90.00 \\
Nova Pro (\textasciitilde90B) & 88.00 & 37.00 & 88.00 & 88.00 \\
Llama 3.3 70B (70B) & 93.00 & 37.00 & 77.00 & 93.00 \\
\midrule
\textbf{CDMizer with Qwen3 (30.5B)} & \textbf{88.81} & \textbf{58.31} & \textbf{46.22} & \textbf{82.37} \\
\textbf{Rank (out of 9 models)} & \textbf{7\textsuperscript{th}} & \textbf{1\textsuperscript{st}} & \textbf{7\textsuperscript{th}} & \textbf{9\textsuperscript{th}} \\
\bottomrule
\end{tabular}
\end{table*}

\subsection{Template Creation and Scope}
We focused on the five critical CSA clauses identified in the ISDA benchmarking study~\cite{csa_benchmark}: \textbf{Base Currency, Eligible Currency, Rounding, Minimum Transfer Amount (MTA), and Threshold}.

As detailed in the methodology of the CDMizer framework~\cite{mridul2025ai4contractsllmragpowered}, template creation is a deterministic process (Algorithm 1 in the cited work). It involves recursively traversing the full CDM schema and pruning all fields not relevant to the target clauses, resulting in a minimal, syntactically correct JSON structure. This process yielded four unique templates: one for the combined \textbf{Base Currency} and \textbf{Eligible Currency} fields, and one each for \textbf{Rounding}, \textbf{MTA}, and \textbf{Threshold}. This template-driven structure guarantees that the output JSON will always be compliant with the defined CDM schema, thus achieving \textbf{$100\%$} syntactical correctness and schema adherence, an advantage over direct generation methods.

\subsection{LLM and Inference Setup}
We selected the \textbf{Qwen3-30B} LLM for this study. This model represents a highly capable, modern open-source architecture with $30.5$ billion total parameters (and approximately $3.3$ billion active parameters per inference), classifying it as a \textit{significantly smaller} model compared to the proprietary SOTA models used in the ISDA benchmark (e.g., those estimated to be over $100$ billion parameters). Our primary goal is to assess the utility of this highly accessible model through the CDMizer framework for this complex, domain-specific task.

We tested two configurations for each clause:
\begin{enumerate}
    \item \textbf{Without RAG:} The LLM receives only the clause-specific template, the schema definition, and the natural language contract text.
    \item \textbf{With RAG:} The LLM receives the prompt augmented with contextual examples retrieved from a RAG knowledge base.
\end{enumerate}
For RAG implementation, the knowledge base consisted of the $60$ ground truth CSA examples. We employed a \textit{leave-one-out} approach: 
when processing a given CSA, we excluded it from the retrieval pool and instead allowed the RAG system to draw contextual examples only from the remaining $59$. In other words, each contract was evaluated as if it were a ``new'' unseen document, while the model could still learn patterns from similar but different CSAs. This ensured that no CSA was ever evaluated against its own ground truth, preserving fairness and preventing information leakage.

\subsection{Dataset and Evaluation}
\paragraph{\textbf{Dataset:}} We used a dataset of $60$ real-world CSA samples, which are the same source contracts leveraged in the ISDA benchmarking study~\cite{csa_benchmark}. 
Using this dataset ensures direct comparability with industry-standard benchmarks and provides a realistic assessment of performance on authentic legal documents.
Note that the \textbf{Threshold} clause was applicable to $37$ of the samples, resulting in a smaller test count for that clause. 

\paragraph{\textbf{Evaluation Metrics:}} The accuracy of the generated CDM outputs was assessed through manual, human evaluation, mirroring the ISDA benchmark's methodology to ensure consistency and relevance. We manually compared the generated CDM JSON against the ground truth for each clause and assigned a final score on a scale of $0$ to $100$. The ground truth data for these CSA samples was human-annotated by ISDA experts, making it a reliable source of reference for evaluating the performance of our model. The evaluation score reflects:
\begin{itemize}
    \item Correct schema usage (which CDMizer already guarantees).
    \item Correct extraction of the target information from the contract text.
    \item Correct placement and representation of the extracted values (e.g., currency codes, numeric values, party names) within the structured CDM fields.
\end{itemize}

To illustrate the nature of the dataset, following is an example of an MTA clause and it's corresponding CDM representation~\cite{csa_benchmark}.

\begin{tcolorbox}[title=An MTA Clause Excerpt from a CSA, coltitle=white, fonttitle=\bfseries, sharp corners=southwest, enhanced]

Paragraph 12. Definitions 
“Minimum Transfer Amount” means, with respect to a party, the amount specified as such for that party in Paragraph 13; if no amount is specified, zero.

Paragraph 13(vii) Minimum Transfer Amount

(A) “Minimum Transfer Amount” means with respect to Party A: US Dollars 5,000,000. “Minimum Transfer Amount” means with respect to Party B: US Dollars 5,000,000.

\end{tcolorbox}

\begin{tcolorbox}[title= CDM JSON Representation for the MTA Clause Excerpt, coltitle=white, fonttitle=\bfseries, sharp corners=southwest, enhanced]
\ttfamily\small
\begin{verbatim}
"agreementTerms": {
    "agreement": {
      "creditSupportAgreementElections": {
        "minimumTransferAmount": [
          {
            "mtaType": {
              "fixedAmount": {
                "amount": 5000000,
                "currency": "USD",
                "party": "PARTY_1"
              }
            }
          },
          {
            "mtaType": {
              "fixedAmount": {
                "amount": 5000000,
                "currency": "USD",
                "party": "PARTY_2"
              }
            }
          }
        ]
      }
    }
}
\end{verbatim}
\end{tcolorbox}

\section{Results and Discussion}

The results of our evaluation are benchmarked against models tested by ISDA’s benchmarking work~\cite{csa_benchmark}, ranging from large proprietary models (e.g., GPT-4o, $\sim$200B) to open-source alternatives (e.g., Llama 3.3, 70B). The full comparison is detailed in Table \ref{tab:rag_comparison} and Table \ref{tab:baseline_comparison}.

Despite being significantly smaller, the Qwen3-30B model demonstrated competitive performance in CSA-to-CDM conversion. A key strength of the CDMizer framework is its template-driven approach, which guarantees strict schema adherence. This was particularly advantageous when using a smaller model like Qwen3-30B, as it ensured accurate, structured outputs, which are often challenging for direct LLM generation, especially with complex documents like CSAs.

A noteworthy result was seen in the MTA clause: when RAG was not used, Qwen3-30B ranked first in performance. This was primarily due to the CDMizer framework's strict schema adherence, which ensured the correct structure of the output, boosting its score even when compared to larger models. For more standardized clauses like Base and Eligible Currency and Rounding, Qwen3-30B also performed strongly, although larger models like GPT-4o and Claude achieved higher scores across most of the clauses.

When RAG was incorporated, it enhanced the performance of all models, with the most significant improvement observed in Qwen3-30B. RAG helped increase semantic coverage and extraction accuracy, especially for more complex clauses. However, even without RAG, the Qwen3-30B model was able to achieve competitive results, showcasing the potential of smaller open-source models for contract digitization.

Overall, CDMizer with Qwen3-30B, while not ranking at the top in all aspects, achieved scores close to the highest-performing models, demonstrating its ability to generate syntactically correct and semantically meaningful CDM representations on par with larger models, all while remaining computationally efficient. This positions CDMizer as a resource-efficient solution for CSA-to-CDM conversion, demonstrating the framework's broader applicability beyond its initial use for derivatives contracts.

\section{Conclusion}

In this work, we extended the CDMizer framework~\cite{mridul2025ai4contractsllmragpowered} to address the CSA-to-CDM conversion, utilizing the Qwen3-30B model and benchmarking its performance against the ISDA standard. Despite its smaller size, Qwen3-30B with CDMizer showed competitive results, especially when integrated with RAG, achieving accuracy levels close to those of larger models. This highlights the potential of resource-efficient solutions for automating contract digitization at scale without compromising on performance. Future work could focus on expanding CDMizer to handle a broader range of legal documents, enhancing the framework's generalization capabilities. Additionally, exploring advanced fine-tuning techniques for domain-specific tasks and incorporating additional contextual knowledge could enhance the framework's robustness, further improving the accuracy and reliability of the conversion process.

\section*{Acknowledgments}
{We acknowledge the support from NSF IUCRC CRAFT center research grant (CRAFT Grant \#22018) for this research. The opinions expressed in this publication do not necessarily represent the views of NSF IUCRC CRAFT. We are also grateful for the advice and resources from our CRAFT Industry Board members, and David Lee from ISDA in shaping this work.}

\balance
\bibliographystyle{ACM-Reference-Format}
\bibliography{references}

\end{document}